\documentstyle[preprint,prd,eqsecnum,aps]{revtex}
\tighten
\begin{document}
\draft
\title {The canonical transformations %\\
of the dynamical multiparameter systems %\\
as recurrence relations %\\
for the models on the grating
}
\vskip 0.5cm
\author{V.D. Gladush\thanks{E-mail: gladush@ff.dsu.dp.ua}
\and
 A.V. Shcherbakov}
\address{ Dniepropetrovsk State University\\
49050 Dniepropetrovsk, st. Naukova, 13}
\date{\today}
\maketitle
\begin{abstract}
The theory of recurrence relations of linear multi-component and
multi-parameter
systems on the basis of the canonical transformations theory
of the dynamical systems'\ sets is constructed. The parameters of the
grating's knots are defined from the condition of the invariance of the model under
shifts along the grating. The connection with a zero curvature representation
for models on the grating is installed. The examples of two- and
three-parameter systems described by the hypergeometric functions $M(\alpha,
\beta, t)$ and $M(\alpha, \beta, \xi, t)$ are considered in details. The
canonical recurrence relations increasing and decreasing parameters
$\{ \alpha, \beta, \xi \}$ for solutions of the corresponding equations are
constructed.
\end{abstract}
%\pacs{PACS number(s): 02.40.Ky, 04.20.Cv, 04.20.-q}
\narrowtext
\section{Introduction}
 Many equations of the theoretical physics especially of quantum
mechanics
are linear. In this connection there is a possibility of constructing
solutions of these equations searching the canonical transformations of the
linear dynamical systems (but this canonical transformations are different
from those of quantum theory). These transformations can be formulated in
terms of the creation and annihilation operators for some ``quantum numbers''
of the corresponding quantities that conserve their values. The indicated
possibility was
considered in \cite{1}, where the method of constructing the recurrence
relation for searching for the eigenvalues and eigenfunctions of linear
operators is proposed.
This method is restricted by the case of one-dimensional systems
described by the Lagrangian containing the spectral parameters in a potential
part only.

      In the present paper the generalization of this method for the case of
the multidimensional systems and systems with the Lagrangian containing
spectral parameters in a kinetic part too (for example, for the radial
equation in hydrogen-like atom theory) is constructed.

      The physical meaning of the considered canonical transformations
for the classical theory differs from that for the quantum theory. The
general point
is that these transformations can be considered as translations on the
multidimensional grating in the phase space. In the classical theory the
parameters of the grating's knots define the different configurations
of considered class of dynamical systems (for example, the set of
mechanical oscillators with different masses and spring rigidities).
Conditions of the invariance under the canonical transformations along the
grating define the admissible grating parameters and, therefore,
solutions and a spectrum of oscillations.

      In quantum theory the parameters of the knots give the possible states
of the same quantum system. In its turn the translation along the
grating is defined by the creation and annihilation operators stipulating
the transition from one state to another.

      The canonical recurrence transformations obtained in this way  are the
analogies of the known Beclund's \cite{2} transformations for
finite-dimensional dynamical systems. Note that they do not
coincide by their form with the standard recurrence relations cited in handbooks.
The difference between them is explained by the way of their construction.
The main here is that the canonical recurrence relations conserve the
symplectic structure of the corresponding dynamical systems in contrast with
the standard ones. That is why from the standpoint of the theory of dynamical
systems the canonical recurrence relations have a certain advantage.
This relations can
be reduced to the standard ones by the corresponding normalization of
eigenfunctions.

\section{A canonical map}
      Let us consider the sets ${\cal D} = \{q,p,H\}$ and $\tilde{\cal D}=
 \{ \tilde{q}, \tilde{p}, \tilde{H}\}$ of the $s$-dimensional dynamical systems
with the generalized coordinates $q=q(t)\equiv\{q^{1}(t),..,q^{s}(t)\}$,
$\tilde{q}=\tilde{q}(t)\equiv\{\tilde{q}^{1}(t),\ldots,\tilde{q}^{s}(t)\}$, momenta
$p=p(t)\equiv\{p_{1}(t),\ldots,p_{s}(t)\}$,
$\tilde{p}=\tilde{p}(t)\equiv\{\tilde{p}_{1}(t),\ldots,\tilde{p}_{s}(t)\}$
and Hamiltonians $H=H(p,q,t)$, $\tilde{H}=\tilde{H}(\tilde{p},\tilde{q},t)$.

      We shall call the mapping ${\cal F}:{\cal D}\rightarrow\tilde{\cal D}$
the canonical one, if it is defined by the set ${\cal F}=\{F\}$ of generating
functions $F=F(q,\tilde{q},t)$, so that for an arbitrary system
$(q,p,H)\in{\cal D}$ and for some function $F\in{\cal F}$ there is the
system
$(\tilde{q},\tilde{p},\tilde{H})\in\tilde{\cal D}$, for which the relations
\begin{equation}\label{1}
   p_{i}=\frac{\partial F}{\partial q^{i}},\quad
   \tilde{p}_{i}=-\frac{\partial F}{\partial\tilde{q}^{i}},\quad
   \tilde{H}=H+\frac{\partial F}{\partial t}\quad (i,j,k,\ldots=1,2,\ldots,s),
\end{equation}
are realized.

      If ${\cal D}=\tilde{\cal D}$, we shall say that given set of
dynamical systems ${\cal D}$ is invariant under the canonical transformation
$F$.

        Let  ${\cal{D}}$ be a parametrized class of dynamical systems
described by the Lagrangian
\begin{equation}\label{2}
   L=L(q,\dot{q},t,\lambda,\mu_{a})=\frac{1}{2}\left[g_{ij} \dot{q}^{i}
   \dot{q}^{j}-
   (U_{ij}-\lambda g_{ij}) q^{i} q^{j}\right],
\end{equation}
where $ \dot{q}=dq/dt$, $g_{ij}=g_{ij}(\mu_{a},t)$, $U_{ij}=U_{ij}(\mu_{a},t)$
are some functions, $\det||g_{ij}||\not=0$, $\{\lambda,\mu_{a}\}\in D $,
$D$ is some set of parameters, $a=1,\ldots,d$. The ordinary summation
convention works here and on.

      If we write the Lagrange-Euler equations for the system (\ref{2})
in the form
\begin{equation}\label{3}
   \hat{P}^{l}_{k}q^{k}\equiv g^{li}\left(-\frac{d}{dt} g_{ik}
   \frac{d}{dt} + U_{ik}\right) q^{k}=\lambda q^{l}
\end{equation}
we shall make a conclusion that $\lambda$ is a spectral parameter of the
eigenvalue problem of the $d$-parametric set of the linear operators
$\hat{P}^{l}_{k}(\mu_{a},t)$. For the Hamiltonians of the set (\ref{2}) we
have
\begin{equation}\label{4}
   H(p,q,t,\lambda,\mu_{a})=\frac{1}{2} g^{ij}p_{i}p_{j} -\frac{1}{2}(U_{ij}-
   \lambda g_{ij})q^{i}q^{j},
\end{equation}
where
\begin{equation}\label{5}
  p_{i}=g_{ij} \dot{q}^{j},\quad (g^{ik} g_{kj}=\delta^{i}_{j}).
\end{equation}

        The condition of the invariance of the set ${\cal{D}}$ of dynamical
systems under the canonical transformations $\cal{F}$ is
\begin{equation}\label{6}
  \tilde{H}(\tilde{p},\tilde{q},t,\tilde{\lambda},\tilde{\mu}_{a})=
  H(\tilde{p},\tilde{q},t,\tilde{\lambda},\tilde{\mu}_{a}),\quad
  \forall \{\tilde{\lambda},\tilde{\mu}_{a}\}\in D.
\end{equation}

        Under sequential actions of the series of the canonical
transformations a sequence of the systems of the type (\ref{2}) and the
parameters $\{\lambda_{m},\mu_{n_{a}}| m,n_{a}=1,2,\ldots\}\in D$ appears.
Introduce the collective indices $K=(m,n_{a})$,
$\tilde{K}=(\tilde{m},\tilde{n}_{a})$ and designate
$(\lambda_{m},\mu_{n_{a}})\equiv\sigma_{K}$,
$(\lambda_{\tilde{m}},\mu_{\tilde{n_{a}}} )\equiv\sigma_{\tilde{K}}$.
Let the corresponding coordinates and momenta be $q_{K},q_{\tilde{K}}$ and
$p_{K},p_{\tilde{K}}$. Further,
$F_{K\tilde{K}}=F_{K\tilde{K}}(q_{K},q_{\tilde{K}},t)$ be the generating function
of the canonical transformation $F$: $\cal{D}\rightarrow\cal{D}$, so that
$\left\{q_{K},p_{K},H(p_{K},q_{K},t,\sigma_{K})\right\}$ $\rightarrow$
$\left\{q_{\tilde K},p_{\tilde K},H(p_{\tilde{K}},q_{\tilde{K}},t,\sigma_{\tilde{K}})
\right\}$.
Then the conditions (\ref{1}), (\ref{6}) give the following equation for the
generating function
\begin{equation}\label{7}
 \frac{\partial F_{K \tilde{K}} }{\partial t}+H \left(\frac
 {\partial F_{K \tilde{K}} } {\partial q_{K}},q_{K},\sigma_{K},t \right)=
 H \left( -\frac{\partial F_{K \tilde{K}} }{\partial q_{\tilde{K}} },
 q_{\tilde{K}},\sigma_{\tilde{K}},t \right).
\end{equation}
Further we shall write the indices $K$ and $\tilde{K}$ for the generating
functions only, and for an arbitrary physical variable $f$ we shall write
$f$ or $\tilde{f}$ instead of $f_{K}$ or $f_{\tilde{K}}$ respectively. Then it is
possible to rewrite the condition of the invariance (\ref{7}) for the
system (\ref{2}) with the Hamiltonian (\ref{4}) in the following form:
\begin{equation}\label{8}
  2\frac{\partial F_{K \tilde{K}} }{\partial t}+g^{ij}\frac{ \partial
  F_{K \tilde{K}} }{\partial q^{i}}
  \frac{\partial F_{K \tilde{K}} }{\partial q^{j}}-\tilde{g}^{ij}
  \frac{\partial F_{K \tilde{K}} }{\partial \tilde{q}^{i}}
  \frac{\partial F_{K \tilde{K}} }{\partial \tilde{q}^{j} }=
  (U_{ij}-\lambda g_{ij})q^{i}q^{j}-(\tilde{U}_{ij}-
  \tilde{\lambda}\tilde{g}_{ij})\tilde{q}^{i}\tilde{q}^{j}.
\end{equation}

\section{ A recurrence relations as a canonical map}
Let us consider the problem of searching for the solutions of
Eq.(\ref{8}). The canonical map, that converts the linear dynamical
systems into linear ones, must be linear. That is why the generating
function $F_{K \tilde{K}}$ must be a quadratic function of the generalized
coordinates $q,\tilde{q}$. Therefore we search for $F_{K \tilde{K}}$ in the form
\begin{equation}\label{9}
 F_{K \tilde{K}}=\frac{1}{2}\left( 2\gamma_{ij} q^{i}\tilde{q}^{j}-b_{ij}q^{i}q^{j}-
  c_{ij}\tilde{q}^{i}\tilde{q}^{j} \right),
\end{equation}
where $\gamma_{ij},b_{ij},c_{ij}$ are some functions. Substituting
Eq.(\ref{9}) for Eq.(\ref{8}) and equating the coefficients at
$q^{i}q^{j}$, $q^{i}\tilde{q}^{j}$, $\tilde{q}^{i}\tilde{q}^{j}$, we obtain
\begin{eqnarray}\label{10}
 2\dot\gamma_{rs}-g^{ij}(\gamma_{ir}b_{js}+\gamma_{jr}b_{is})=
           -\tilde{g}^{ij}(\gamma_{is}c_{jr}+\gamma_{js}b_{ir}), \\
 -\dot b_{rs}+g^{ij}b_{ir}b_{js}+\lambda g_{rs}-U_{rs}=
            \tilde{g}^{ij}\gamma_{ir}\gamma_{js},
\end{eqnarray}
\begin{equation}\label{12}
-\dot c_{rs}+g^{ij}\gamma_{ir}\gamma_{js}=\tilde{g}^{ij}c_{ir}c_{js}-
           \tilde{U}_{rs}+\tilde{\lambda}\tilde{g}_{rs}.
\end{equation}
If we find one of the particular solutions $\gamma_{ij},b_{ij},c_{ij}$, then,
using Eqs.(\ref{1}), (\ref{5}), (\ref{9}), it will be easy to construct the
binomial recurrence relations for the eigenfunctions $q(t)$ and $\tilde{q}(t)$
of the set of operators $\hat{P}^{l}_{k}$
\begin{equation}\label{13}
 \left( b_{ij}+g_{ij}\frac{d}{dt} \right)q^{j}=\gamma_{ij}{\tilde q}^{j},
\end{equation}
\begin{equation}\label{14}
 \left( c_{ij}-{\tilde g}_{ij}\frac{d}{dt} \right){\tilde q}^{j}=\gamma_{ij}q^{j}.
\end{equation}
The obtained relations, increasing and decreasing $K$, are the analogies of
the well-known Beclund's transformations \cite{2} for finite-dimensional
systems. They are also the generalization of the known relations of the
factorization method \cite{3}, \cite{4}. It means, that Eqs.(\ref{13}),
(\ref{14}) install the correlation between the canonical transformation
method and factorization one.

      The algebraic trinomial recurrence relations can be obtained
considering the sum of the generating functions of the two sequential
transformations

\begin{equation}\label{15}
   F_{K \tilde{K} \tilde{\tilde{K}}}=F_{K \tilde{K}}+F_{\tilde{K}\tilde{\tilde{K}}}.
\end{equation}

      According to Eq.(\ref{1}) we have
$\partial F_{K \tilde{K} \tilde{\tilde{K}}}/\partial \tilde{q} =0$, from which the
linear algebraic recurrence relation
\begin{equation}\label{16}
 \gamma_{ij} q^{j}-(c_{ij}+\tilde{b}_{ij}) \tilde{q}^{j} + \tilde{\gamma}_{ij}
\tilde{\tilde q}^{j}=0,
\end{equation}
follows. Here $\tilde{b}_{ij},\tilde{c}_{ij},\tilde{\gamma}_{ij}$ are the
coefficients of a quadratic form of the type (\ref{9}) for the generating
function $F_{\tilde{K} \tilde{\tilde{K}}}$.

      The generating function $F_{K \tilde{\tilde{K}}}$ of the canonical
transformation $(q,p)\rightarrow(\tilde{\tilde{q}},\tilde{\tilde{p}})$, which is a
composition of the canonical transformations $(q,p)\rightarrow(\tilde{q},\tilde{p})$ and
$(\tilde{q},\tilde{p})\rightarrow(\tilde{\tilde{q}},\tilde{\tilde{p}})$, can be constructed
excluding the intermediate state $\tilde{q}$ from Eq.(\ref{15})
by means of Eq.(\ref{16}). It makes possible to simplify the procedure of
searching for the generating function of the transformation $K\rightarrow\tilde{K}$
decomposing it into more simple steps of the calculating the elementary
generating functions of the transformations:
$K=\{m,n_{1},\ldots\}\rightarrow\{m+1,n_{1},\ldots\}\rightarrow\{m+1,n_{1}+1,\ldots\}\rightarrow \cdots
\rightarrow\tilde{K}$,
and after this to construct their composition.

        From the standpoint of classical mechanics we consider here
the canonical transformation theory of an infinite-dimensional system with
the action
$$ S=\frac{1}{2} \int dt\sum_{K}
   \left\{g_{ij} \dot{q}^{i}_{K}
\dot{q}^{j}_{K}+(U_{ij}-\lambda_{m} g_{ij})q^{i}_{K} q^{j}_{K}\right\},$$
describing a multicomponent model on a $(d+1)$-dimensional grating. The values
of the parameters  $\sigma_{K}=\{\lambda_{m},\mu_{n_{1}},\ldots,\mu_{n_{d}}\}$
that define grating knots are determined from the condition of an invariance
of the dynamical system under grating translations induced by the canonical
transformations of the system.

        One more interesting possibility connected with the method of the
inverse scattering problem is shown below by the examples of more simple
dynamical systems.

        For the particular case of the set of $d$-component three-parametrical
dynamical systems  $(s=1)$ let us denote:
$g_{11} \equiv m(\nu,t)$, $U_{11}\equiv m(\nu,t)U(\mu,t)$,
where $\mu,\nu$ are parameters. Write the Lagrangian (\ref{2}) in
the form
\begin{equation}\label{17}
 L= \frac{m(\nu,t)}{2} \left [\dot{q}^2+(U(\mu,t)-\lambda)q^2 \right ].
\end{equation}
It is convenient to find the generating function $F_{K \tilde K}$ in such a form
\begin{equation}\label{18}
  F_{K \tilde K}=mS,\quad S=\frac{1}{2a}(2\gamma q\tilde q-bq^2-c\tilde {q}^2),
\end{equation}
where $\gamma$ is some unknown constant, $a,b,c$ are desired functions of $t$.
The conditions (\ref{1}) may be rewritten in the form
\begin{equation}\label{19}
 \dot{q}=\frac{\partial S}{\partial q},\quad
 \dot{\tilde{q}}=-\frac{m}{\tilde{m}}\frac{\partial S}{\partial \tilde{q}},\quad
 \tilde{H}=H+\frac{\partial }{\partial t}(mS).
\end{equation}
The recurrence relations
\begin{equation}\label{20}
 \left(b+a\frac{d}{dt} \right)q=\gamma \tilde{q},\quad
 \left(c-a\frac{m}{\tilde{m}}\frac{d}{dt} \right)\tilde{q}=\gamma q
\end{equation}
and equation for the function $S$
\begin{equation}\label{21}
 \frac{2}{m} \frac{\partial} {\partial t}(mS)+\left (\frac{\partial S}
 {\partial q}\right)^{2}- \frac{m}{\tilde m} \left (\frac{\partial S}
 {\partial t}\right)^{2}-(\lambda-U)q^{2}-\frac{\tilde m}{m}(\tilde{\lambda}-
 \tilde U)\tilde{q}^2=0.
\end{equation}
follow from Eq.(\ref{6}), (\ref{19}).
Eqs. (\ref{10})--(\ref{12}) for determining  $a,b,c,\gamma$ acquire the form
\begin{equation}\label{22}
   \dot{a}=\frac{\dot m}{m}a+\frac{m}{\tilde m}c-b,
\end{equation}
\begin{equation}\label{23}
 \frac{m}{\tilde m} \dot c+\dot b=
 (\lambda-\tilde{\lambda}+{\tilde U}-U)a,
\end{equation}
\begin{equation}\label{24}
 a\dot b =(\lambda-U) a^{2}+\frac{m}{\tilde m} (bc-\gamma^{2}).
\end{equation}
The system of equations (\ref{22})--(\ref{23}) is linear
with respect to the unknown
functions $a,b,c$ and does not contain the constant $\gamma $, while
the equation (\ref{24}) is quadratic and contains $ \gamma $. The procedure
of the searching for the solutions of the system
(\ref{22})--(\ref{24}) consists in the finding a more simple particular
solution of the homogeneous linear undetermined system (\ref{22})-
(\ref{23}) for which the quadratic equation (\ref{24}) is satisfied at the
some unknown parameters $\{\tilde{\lambda},\tilde{\mu},\tilde{\nu}\}$. Then
the constant $\gamma$ will be determined.

        The algebraic trinomial recurrence relations (without derivatives)
and their compositions can be found by the way analogous to obtaining
Eq.(\ref{16}) (see \cite{1} also). As a result we shall have
\begin{equation}\label{25}
\frac{ \gamma_{K\,K+1}}{a_{K}}q_{K}-\left(\frac{c_{K+1\,K+1}}{a_{K}}+
\frac{b_{K+1\,K+1}}{a_{K+1}}\right)q_{K+1}+
\frac{\gamma_{K+1\,K+2}}{a_{K+1}}q_{K+2}=0,
 \end{equation}
where $K$ is the collective index of the totality of the parameters
$ \{\lambda_{m},\mu_{n},\nu_{l}\}$, i.e. if $K=\{m,n,l\}$ then $K+1$ means
that one of the indices $m,n,l$ is increased by one. The coefficients
in Eq.(\ref{25}) correspond to that of the generating functions $S_{K\,K+1}$
and $S_{K+1\,K+2}$ of the type (\ref{18}) of the transformations
$q_{K}\rightarrow q_{K+1}$ and $q_{K+1}\rightarrow q_{K+2}$.

      Writing the similar equations for the sequence of the transformations
$ q_{K+1}\rightarrow q_{K+2}\rightarrow q_{K+3} $ and excluding the coordinate
$q_{K+2}$ by means of Eq.(\ref{25}), we obtain the two relations, which can be
written in the matrix form
\begin{equation}\label{26}
  Q_{K+2}=B_{K}Q_{K},
\end{equation}
where
\begin{equation}\label{27}
 Q_{K}=\left(\begin{array}{c}q_{K}\\q_{K+1}\end{array}\right),\quad
 Q_{K+2}=\left
 (\begin{array}{c}q_{K+2}\\q_{K+3}\end{array}\right),
\end{equation}
and $B_{K}$ is a $2\times 2$ matrix, which is expressed through the
coefficients of the quadratic forms $\{S_{K\,K+1},\ldots, S_{K+2\,K+3}\}$ of the
above sequence of the transformations; its explicit form is not
indicated here because of its inconvenience.

        From the other hand the differential equation (\ref{20}) can be
rewritten in these terms as
\begin{equation}\label{28}
 \frac{dQ_{K}}{dt}=A_{K}Q_{K},
\end{equation}
where
\begin{equation}\label{29}
A_{K}=\frac{1}{a_{\scriptscriptstyle K}m_{\scriptscriptstyle K+1}}
\left(\begin{array}{cc}-m_{\scriptscriptstyle K+1}b_{\scriptscriptstyle K\,K}
&m_{\scriptscriptstyle K+1}\gamma_{\scriptscriptstyle K\, K+1}\\
-m_{\scriptscriptstyle K}\gamma_{\scriptscriptstyle K\, K+1}&
m_{\scriptscriptstyle K}c_{\scriptscriptstyle K+1\, K+1}\end{array}\right).
\end{equation}
The conditions of compatibility of Eqs.(\ref{26})--(\ref{27})
\begin{equation}\label{30}
 \frac{dB_{K}}{dt}+B_{K}A_{K}-A_{K+2}B_{K}=0
\end{equation}
are equations for the coefficients $a_{K},b_{K},c_{K},\gamma_{K}$ of the
sequence of the generating functions $S_{K\,K+1}$ and represent zero
curvature conditions for the models on the grating \cite{5}. They are
the conditions of an applicability of the inverse scattering
problem method for integrable models. In our case they are
realized as a consequence of the equations of the type (\ref{22})--(\ref{24}).

       Thus the regular way of the constructing zero curvature conditions,
which can be generalized for nonlinear dynamical integrable systems,
follows from our analysis.

\section{The confluent hypergeometric equation}
Many equations of quantum mechanics under the separation of
variables and after isolation of an angular part lead to the second
order ordinary differential equations, which can be reduced to the
hypergeometric (or confluent hypergeometric) equation. In this connection
it is important to consider the canonical transformation theory of the
equations of such a type in general case.

      Let us consider construction of the canonical recurrence relations
for the confluent hypergeometric equation
\begin{equation}\label{31}
 t\ddot{q}+(\beta-t)\dot{q}-\alpha q=0,
\end{equation}
where $\alpha,\beta$ are some parameters. It is the Lagrange-Euler equation
for a dynamical system with the Lagrangian
\begin{equation}\label{32}
 L=\frac{1}{2}t^{\beta}e^{-t}(\dot{q}^2+\frac{\alpha}{t}q^{2}).
\end{equation}
Comparing Eq.(\ref{32}) with Eq.(\ref{17}), we can conclude that
$$\lambda=0,\quad \nu=\beta,\quad \mu=\alpha,\quad
 U(\alpha,t)=\frac{\alpha}{t},\quad m(\beta,t)=t^{\beta}e^{-t}.$$
Therefore the system (\ref{22})--(\ref{24}) has the form
\begin{equation}\label{33}
 \dot{a}=\frac{\dot m }{m} a+\frac{m}{\tilde{m}}c-b,\quad
 \frac{m}{\tilde{m}}\dot{c}+\dot{b}=\frac{\tilde{\alpha}-\alpha}{t}a
\end{equation}
\begin{equation}\label{34}
 a\dot{b}=-\frac{\alpha}{t}a^{2}+\frac{m}{\tilde{m}}(bc-\gamma^{2})
\end{equation}
According to the above-mentioned, we shall find the generating functions of
the elementary canonical transformations
a) $\{\alpha,\beta\} \rightarrow \{ \tilde{\alpha},\beta\}$, b) $\{\alpha,\beta\}
\rightarrow \{\alpha,\tilde{\beta}\}$ only. Besides note that Eqs.(\ref{33})--(\ref{34})
can be written as equations with rational on $t$ coefficients. Therefore it
is naturally to find the solution in a class of rational functions.

\subsection{A canonical transformation
$\{\alpha,\beta\}\rightarrow\{\tilde{\alpha},\beta\}$}
%      a) First let us consider the transformation $\{\alpha,\beta\}\rightarrow\{\tilde{\alpha},\beta\}$.
In this case
$\beta=\tilde{\beta}$, $m=\tilde{m}$ and the system (\ref{33})--(\ref{34})
acquires the form
\begin{equation}\label{35}
  \dot{a}  =(\frac{\beta}{t}-1)a+c-b,\quad
 \dot{c}+\dot{b}=\frac{\tilde{\alpha}-\alpha}{t}a,
\end{equation}
\begin{equation}\label{36}
 a\dot{b} =bc-\gamma^{2}-a^{2}\frac{\alpha}{t}.
\end{equation}
We shall obtain the simplest rational solution of the undetermined system (\ref{35})
by assumption $a=t$. From (\ref{35}) we shall find
\begin{equation}\label{37}
 b=\frac{1}{2}(\tilde{\alpha}-\alpha-1)t+b_0,\quad
 c=\frac{1}{2}(\tilde{\alpha}-\alpha+1)t+b_0-\beta+1,
\end{equation}
where $b_0$ is the integration constant. Substituting Eq.(\ref{37}) for
Eq.(\ref{36}) and equating the coefficients at the same powers of $t$, one
obtains
\begin{equation}\label{38}
 b_0=\alpha,\quad \tilde{\alpha}=\alpha+1,\quad
 \gamma=\sqrt{\alpha(\alpha-\beta+1)}
\end{equation}
We have restricted ourselves here by the positive value of the radical. As a
result  we have
\begin{equation}\label{39}
 F_{\alpha,\beta \mid \alpha+1,\beta}=\frac{1}{2}t^{\beta-1}e^{-t}
 \left( 2 \sqrt{\alpha(\alpha-\beta+1)} q\tilde{q}-\alpha q^{2}-
 (t+\alpha-\beta+1)\tilde{q}^{2} \right)
\end{equation}
for the generating function (\ref{18}).
The canonical recurrence transformations (\ref{20}) have the form
\begin{equation}\label{40}
 \left( \alpha+t \frac{d}{dt} \right) q= \sqrt{\alpha ( \alpha-\beta+1)}\tilde q ,\quad
 \left( t+\alpha-\beta+1-t\frac{d}{dt}\right) \tilde q=\sqrt{\alpha(\alpha-\beta+1)}q.
\end{equation}
To determine the connection with the standard recurrence relations and thus
with the confluent hypergeometric function $M(\alpha,\beta,t)$, we shall make
the following substitutions
\begin{equation}\label{41}
 q=y(\alpha)\sqrt{N(\alpha)},\quad
 \tilde{q}=y(\alpha)\sqrt{N(\alpha+1)},\quad
 y(\alpha) \equiv M(\alpha,\beta,t)
\end{equation}
in Eq.(\ref{40}). From the condition of coincidence of the obtained
recurrence relations with the standard ones \cite{6} (rather with their
consequence)
\begin{equation}\label{42}
 \left( \alpha+t\frac{d}{dt} \right) y(\alpha)=\alpha y(\alpha+1),\quad
 \left( \alpha-\beta+t-1-t\frac{d}{dt} \right) y(\alpha+1)=(\alpha-\beta+1)y(\alpha)
\end{equation}
the functional equation for the normalization multiplier
$N(\alpha)$
\begin{equation}\label{43}
 \alpha N(\alpha)=(\alpha-\beta+1)N(\alpha+1)
\end{equation}
follows. It can be solved by using the gamma-function's property $\Gamma(z+1)=
z\Gamma(z)$ from where the substitutions
\begin{equation}\label{44}
 \alpha=\frac{\Gamma(\alpha+1)}{\Gamma(\alpha)},\quad \alpha-\beta+1=
\frac{\Gamma(\alpha-\beta+2)}{\Gamma(\alpha-\beta+1)}.
\end{equation}
follow. Using these substitutions, equation (\ref{43}) can be reduced to the
form
\begin{equation}\label{45}
 \frac{\Gamma(\alpha-\beta+1)}{\Gamma(\alpha)}N(\alpha)=
 \frac{\Gamma(\alpha-\beta+2)}{\Gamma(\alpha+1)}N(\alpha+1)=
 \frac{\Gamma(\alpha-\beta+3)}{\Gamma(\alpha+2)}N(\alpha+2)=\cdots \, .
\end{equation}
The additional equations arise owing to considering the next
elementary steps by the parameter $\alpha$. From here it is easy to see that
$\Gamma(\alpha-\beta+1)N(\alpha)/\Gamma(\alpha)=C_{1}={\rm const}$. Therefore
$$ N(\alpha)=C_{1} \frac{\Gamma(\alpha)}{\Gamma(\alpha-\beta+1)}.$$

        Thus, we have obtained the ``canonical'' solution of the confluent
hypergeometrical equation in the form
\begin{equation}\label{47}
     q(t)=C_{1} M(\alpha,\beta,t)
     \sqrt{\frac{\Gamma(\alpha)}{\Gamma(\alpha-\beta+1)}}.
\end{equation}

\subsection{A canonical transformation $\{\alpha,\beta\}
 \rightarrow \{\alpha,\tilde{\beta}\}$}
%        b) Now let us consider the elementary transformation
%$\{\alpha,\beta\}\rightarrow \{\alpha,\tilde{\beta} \}$.
In this case $\alpha=\tilde{\alpha}$ and the system (\ref{33})--(\ref{34}) can be rewritten
in the form
\begin{equation}\label{49}
    t^{\beta-\tilde{\beta}}c-b=\dot{a}+(1-\frac{\beta}{t})a,\quad
    t^{\beta-\tilde{\beta}}\dot{c}+\dot{b}=0,\\
\end{equation}
\begin{equation}\label{50}
   a \dot{b}+\frac{\alpha}{t}a^{2}+t^{\beta-\tilde{\beta}}(\gamma^{2}-bc)=0.
\end{equation}
We shall obtain the simplest rational solution of the inmohomogeneous
undetermined system
(\ref{49}) by taking $a=1$. Then a particular solution of the obtained
nonhomogeneous system has the form
\begin{equation}\label{51}
 c=\frac{\beta}{\tilde{\beta}-\beta+2}t^{\tilde{\beta}-\beta-1},\quad
 b=\frac{\tilde{\beta}-\beta-1}{\tilde{\beta}-\beta+2}~\frac{\beta}{t}\quad .
\end{equation}
From the quadratic equation (\ref{50}) we find
\begin{equation}\label{52}
 \tilde{\beta}=\beta+1,\quad \gamma=\sqrt{\beta-\alpha}.
\end{equation}
For the generating function and canonical recurrence transformations we have
\begin{equation}\label{53}
 F_{\alpha,\beta \mid \alpha,\beta+1}=\frac{1}{2}t^{\beta}e^{-t}
 \left(2\sqrt{\beta-\alpha}~q\tilde q+q^2+\beta\tilde q ^2 \right),
\end{equation}
\begin{equation}\label{54}
 \left(-1+\frac{d}{dt} \right) q=\sqrt{\beta-\alpha}~\tilde q,\quad
 -\left( \beta+t\frac{d}{dt} \right) \tilde q=\sqrt{\beta-\alpha}~q.
\end{equation}
The recurrence relations increasing and decreasing $\beta$ that follow
from the recurrence relations of the handbook \cite{6}
\begin{equation}\label{55}
 \beta \left(-1+\frac{d}{dt} \right) y(\beta)=(\beta-\alpha)y(\beta+1),\quad
 -\left( \beta+t\frac{d}{dt} \right) y(\beta+1)=\beta y(\beta),
\end{equation}
where $y(\beta) \equiv M(\alpha,\beta,t)$, can be obtained from
Eq.(\ref{54}) by the above procedure of the constructing of Eq.
(\ref{42})--(\ref{47}). As a result one has
\begin{equation}\label{56}
 q(t)=C_{2}\frac{\Gamma(\beta)}{\sqrt{\Gamma(\beta-\alpha)}}M(\alpha,\beta,t).
\end{equation}

\section{The hypergeometric equation}
As it was mentioned, some ``radial equations'' of the quantum
theory are reduced to the hypergeometric-like equations. Therefore
consider a dynamical system described by the hypergeometrical equation
\begin{equation}\label{57}
 t(1-t)\ddot q-((\alpha+\beta+1)t-\zeta)\dot{q}-\alpha\beta q=0,
\end{equation}
where $\alpha,\beta,\zeta$ are some parameters, as another application
of the developed theory. It is the Lagrange-Euler
equation of a dynamical system with the Lagrangian
\begin{equation}\label{58}
 L=\frac{1}{2}t^{\zeta}(1-t)^{\alpha+\beta-\zeta+1}\left(\dot{q}^{2}
 +\frac{\alpha\beta}{t(1-t)}q^{2}\right).
\end{equation}
Comparing Eq.(\ref{58}) with Eqs.(\ref{2}), (\ref{17}) we have:
$\lambda=0,\quad \mu_{1}=\alpha,\quad \mu_{2}=\beta,\quad \mu_{3}=\zeta,$
\begin{equation}\label{a}
     U=\frac{\alpha\beta}{t(1-t)},\quad
     m=t^{\zeta}(1-t)^{\alpha+\beta-\zeta+1}.
\end{equation}
The system (\ref{58}) contains the three parameters $\alpha,\beta,\zeta$.
Therefore it is necessary to find the generating functions of three canonical
transformations. Due to the symmetry between the parameters $\alpha$ and
$\beta$, it is sufficient to search for the generating
functions of the transformations
a) $\{\alpha,\beta,\zeta\}\rightarrow\{\tilde{\alpha},\beta,\zeta\}$
and b) $\{\alpha,\beta,\zeta\}\rightarrow\{\alpha,\beta,\tilde{\zeta}\}$ only.

\subsection{A canonical transformation
 $\{ \alpha,\beta,\zeta \} \rightarrow \{ \tilde{\alpha},\beta,\zeta \} $}
For the elementary transformation
$\{ \alpha,\beta,\zeta \} \rightarrow \{ \tilde{\alpha},\beta,\zeta \} $
the system (\ref{22})--(\ref{24}) can be rewritten the form
\begin{equation}\label{60}
      (1-t)^{\alpha-\tilde{\alpha}}c-b=\dot{a}-\left(\frac{\zeta}{t}
      -\frac{\alpha+\beta- \zeta+1}{1-t}\right)a ,
\end{equation}
\begin{equation}\label{61}
      (1-t)^{\alpha-\tilde{\alpha}}\dot{c}+\dot{b}=\frac{\tilde{\alpha}-\alpha}
      {t(1-t)}\beta a,
\end{equation}
\begin{equation}\label{62}
      a\dot{b}=-\frac{\alpha\beta}{t(1-t)}a^{2}+
      (1-t)^{\alpha-\tilde{\alpha}}(bc-\gamma^{2}).
\end{equation}
Similarly to the first case of the previous example we obtain the
simplest rational solution
of the homogeneous undetermined system (\ref{60})--(\ref{61})
at $a=t$. Then the obtained inhomogeneous system has the following particular
solution
\begin{equation}\label{63}
 b=\alpha-\frac{\tilde{\alpha}-\alpha-1}{\tilde{\alpha}-\alpha-2}\frac{\alpha+\beta
 -\zeta+1}{1-t},\quad
 c=-\left(\beta+\frac{\alpha+\beta-\zeta+1}{\tilde{\alpha}-\alpha-2}
 \frac{1}{1-t}\right)(1-t)^{\tilde{\alpha}-\alpha} .
\end{equation}
In this case the quadratic equation (\ref{62}) is satisfied at
\begin{equation}\label{64}
 \alpha=\tilde{\alpha}+1,\quad \gamma=\sqrt{\alpha(\alpha-\zeta+1)}.
\end{equation}
So for the generating function we have
\begin{equation}\label{65}
 F_{\alpha,\beta,\zeta \mid \alpha+1,\beta,\zeta}=\frac{1}{2}t^{\zeta-1}(1-t)^{
 \alpha+\beta-\zeta+1}\left [2\sqrt{\alpha(\alpha-\zeta+1)}q\tilde{q}-
 \alpha q^{2}-(\alpha-\zeta+1+\beta t)\tilde{q}^2 \right ].
\end{equation}
The corresponding recurrence transformations have the form
\begin{eqnarray} \label{66}
 \left( \alpha+t\frac{d}{dt} \right) q &=& \sqrt{\alpha(\alpha-\zeta+1)}\tilde{q} ,\nonumber \\
 \left( \alpha-\zeta+1-\beta t-t(1-t)\frac{d}{dt} \right) \tilde{q} &=&
 \sqrt{\alpha(\alpha-\zeta +1)}q.
\end{eqnarray}
The standard recurrence transformation \cite{6}
\begin{eqnarray} \label{67}
 \left( \alpha+t\frac{d}{dt} \right) y(\alpha) &=& \alpha y(\alpha+1),\nonumber \\
 \left( \alpha-\zeta+1+\beta t-t(1-t)\frac{d}{dt} \right) \tilde q &=&
 \sqrt{\alpha(\alpha-\zeta+1)}q,
\end{eqnarray}
where $ y(\alpha) \equiv M(\alpha,\beta,\zeta,t) $ --- hypergeometric
function, can be obtained by making such the substitution
\begin{equation}\label{68}
q(t)=A_1 \sqrt{ \frac{\Gamma(\alpha)}{\Gamma(\alpha-\zeta+1)}}\
         M(\alpha,\beta,\zeta,t)
\end{equation}
in Eq.(\ref{66}).

\subsection{A canonical transformation
 $\{\alpha,\beta,\zeta\}\rightarrow\{\alpha,\beta,\tilde{\zeta}\}$}
%  Now let us consider the elementary transformation
%$\{\alpha,\beta,\zeta\}\rightarrow\{\alpha,\beta,\tilde{\zeta}\}$.
In this case the system
(\ref{22})--(\ref{24}) acquires the form
\begin{eqnarray}\label{69}
 t^{\zeta-\tilde \zeta}(1-t)^{\tilde \zeta-\zeta}c-b &=& \dot a-
 \left(\frac{\zeta}{t}-\frac{\alpha+\beta-\zeta+1}{1-t} \right)a,\nonumber\\
 t^{\zeta-\tilde \zeta}(1-t)^{\tilde \zeta-\zeta}\dot c+\dot b &=& 0,
\end{eqnarray}
\begin{equation}\label{70}
 a\dot b=-\frac{\alpha\beta}{t(1-t)}a^{2}+t^{\zeta-\tilde \zeta}
 (1-t)^{\tilde \zeta-\zeta}(b-\gamma^{2}).
\end{equation}
We obtain the simplest solution of the system (\ref{69}) at $a=1-t$.
Then a particular solution (\ref{69}) acquires the form
\begin{equation}\label{71}
  b=\frac{\zeta}{\tilde{\zeta}-\zeta-2}\frac{1-t}{t}+\frac{\zeta}{t}-\alpha-
   \beta,\quad
  c=\frac{\zeta}{\tilde{\zeta}-\zeta-2}t^{\tilde{\zeta}-\zeta-1}(1-t)^{\zeta-\tilde
   {\zeta}+1}.
\end{equation}
The quadratic equation (\ref{70}) leads to the following constraints
\begin{equation}\label{72}
 \tilde{\zeta}=\zeta+1,\quad \gamma=\sqrt{(\zeta-\alpha)(\beta-\zeta)}.
\end{equation}
Therefore for the generating function and the recurrence relations we find
\begin{equation}\label{73}
 F_{\alpha,\beta,\zeta \mid \alpha,\beta,\zeta+1}=\frac{1}{2}t^{\zeta}
(1-t)^{\alpha+\beta-\zeta}\left( 2\sqrt{(\zeta-\alpha)(\beta-\zeta)}q\tilde{q}+
(\alpha+\beta-\zeta)q^{2}+\zeta \tilde{q}^{2}\right),
\end{equation}
\begin{eqnarray}\label{74}
 \left( \zeta-\alpha-\beta+(1-t)\frac{d}{dt} \right) q &=& \sqrt{(\zeta-\alpha)(\beta-\zeta)}
 \tilde q,\nonumber\\
 -\left( \zeta+t\frac{d}{dt} \right)\tilde q &=& \sqrt{(\zeta-\alpha)(\beta-\zeta)}q.
\end{eqnarray}
Making the substitution
\begin{equation}\label{75}
 q(t)= A_{2}
 \sqrt{\frac{\Gamma(\zeta-\alpha)\Gamma(\zeta-\beta)}{\Gamma(\zeta)}}
 M(\alpha,\beta,\zeta,t)
\end{equation}
we reduce Eq.(\ref{74}) to the relations which follow from the
standard recurrence formulae \cite{6} for the hypergeometric function
$M(\alpha,\beta,\zeta,t)$
\begin{eqnarray}\label{76}
 \left( \zeta-\alpha-\beta+(1-t)\frac{d}{dt} \right) y(\zeta) &=& \frac{(\zeta-\alpha)(\zeta-
 \beta)}{\zeta}y(\zeta+1),\nonumber\\
 \left( \zeta+t\frac{d}{dt} \right) y(\zeta+1) &=& \zeta y(\zeta),
\end{eqnarray}
where $y(\zeta) \equiv M(\alpha,\beta,\zeta,t)$.
%\newpage
{
}
\end{document}